\documentclass[12pt]{article}

\newcommand{\argmin}{\mathop{\arg\min}}
\newcommand{\argmax}{\mathop{\arg\max}}
\newcommand{\F}{\text{F}}
\newcommand{\op}{\text{op}}
\newcommand{\ind}{\perp\!\!\!\!\perp}

\newcommand{\dd}{\text{d}}

\providecommand{\inner}[1]{\langle#1\rangle}
\newcommand{\vertiii}[1]{{\left\vert\kern-0.25ex\left\vert\kern-0.25ex\left\vert #1 
    \right\vert\kern-0.25ex\right\vert\kern-0.25ex\right\vert}}

\newcommand{\Dcal}{\mathcal{D}}

\newcommand{\Fcal}{\mathcal{F}}

\newcommand{\Lcal}{\mathcal{L}}

\newcommand{\Xcal}{\mathcal{X}}

\newcommand{\Pscr}{\mathscr{P}}

\newcommand{\EE}{\mathbb{E}}

\newcommand{\II}{\mathbb{I}}

\newcommand{\NN}{\mathbb{N}}

\newcommand{\PP}{\mathbb{P}}

\newcommand{\RR}{\mathbb{R}}

\newcommand{\Bbf}{\mathbf{B}}

\newcommand{\abf}{\mathbf{a}}

\newcommand{\ebf}{\mathbf{e}}

\newcommand{\ubf}{\mathbf{u}}

\newcommand{\xbf}{\mathbf{x}}
\newcommand{\ybf}{\mathbf{y}}
\newcommand{\zbf}{\mathbf{z}}

\newcommand{\thetabf}{{\boldsymbol\theta}}

\newcommand{\xibf}{{\boldsymbol\xi}}

\usepackage{amsthm,amsmath,amssymb,multirow,subfigure,makecell,booktabs,array,url,bm,mathtools,wrapfig,lipsum,mathrsfs,relsize,titling,dsfont,bbm,color,caption, comment,enumitem, geometry, tikz}
\usepackage{algorithm, algpseudocode}
\usepackage{graphicx}
\usepackage{epstopdf,microtype}
\usepackage[export]{adjustbox}[2011/08/13]

\setlist{leftmargin=5mm}
\usepackage[colorlinks,linkcolor=red,citecolor=blue,urlcolor=blue]{hyperref}

\newtheoremstyle{mystyle}{}{}{}{}{\bfseries}{.}{ }
{\thmname{#1}\thmnumber{ #2}\thmnote{ (#3)}}
\theoremstyle{mystyle}
\newtheorem{theorem}{Theorem}[section] 
\newtheorem*{theorem*}{Theorem}

\newtheorem{corollary}[theorem]{Corollary}

\newtheorem{assumption}[theorem]{Assumption}
\newtheorem{remark}[theorem]{Remark}

\usepackage{thm-restate, cleveref}
\usepackage[page,toc,titletoc,title]{appendix}

\makeatletter
\def\keywords{\xdef\@thefnmark{}\@footnotetext}
\makeatother

\usepackage[backend=biber,style=authoryear-comp,sorting=nyt,sortcites=true,giveninits=true,useprefix=true,maxcitenames=2,mincitenames=1,maxbibnames=99,uniquelist=false,uniquename=false,dashed=false,isbn=false,doi=false]{biblatex}
\DeclareDelimFormat{nameyeardelim}{\addcomma\space}
\DeclareDelimFormat[textcite,bib,biblist]{nameyeardelim}{\addspace}
\DeclareDelimFormat{finalnamedelim}{\addspace\&\space}
\DeclareDelimFormat[textcite]{finalnamedelim}{\addspace\bibstring{and}\space}
\DeclareDelimFormat[bib,biblist]{finalnamedelim}{\addcomma\space\&\space}
\DeclareNameAlias{sortname}{family-given}
\DeclareFieldFormat[article,inproceedings,incollection]{title}{#1}
\DeclareFieldFormat[article]{volume}{\mkbibemph{#1}}
\DeclareFieldFormat[article]{number}{\mkbibparens{#1}}
\DeclareFieldFormat[article]{pages}{#1}
\renewbibmacro*{in:}{\ifentrytype{article}{}{\printtext{\bibstring{in}\intitlepunct}}}
\renewbibmacro*{journal+issuetitle}{%
  \usebibmacro{journal}%
  \setunit*{\addcomma\space}%
  \iffieldundef{series}{}{\printfield{series}\setunit{\addspace}}%
  \usebibmacro{volume+number+eid}%
  \setunit{\addspace}\usebibmacro{issue+date}%
  \setunit{\addcolon\space}\usebibmacro{issue}%
  \setunit{\addcomma\space}}
\renewbibmacro*{volume+number+eid}{%
  \printfield{volume}%
  \setunit{}\printfield{number}%
  \setunit{\addcomma\space}\printfield{eid}}
\let\cite\parencite
\AtEveryBibitem{%
  \ifboolexpr{test {\ifentrytype{article}}
    and not test {\iffieldundef{journaltitle}}
    and not test {\iffieldundef{volume}}
    and (not test {\iffieldundef{pages}} or not test {\iffieldundef{eid}})
    and test {\iffieldundef{pubstate}}
    and test {\iffieldundef{entrysubtype}}}
    {\clearfield{url}\clearfield{urlyear}\clearfield{urlmonth}\clearfield{urlday}}
    {}%
}
\usepackage{titletoc}

\newcommand{\worktitle}{Optimal Transport Rank Regression for Robust Multivariate-Response Regression}

\begin{document}

\title{\worktitle}
\author{\large Yizhe Ding$^{\ast}$, Yifan Jiang$^{\dagger}$, Runze Li$^{\ast}$ and Lingzhou Xue$^{\ast}$}
\date{\small $^{\ast}$Department of Statistics, The Pennsylvania State University
\\[0.4em]
$^{\dagger}$Department of Biostatistics, Epidemiology \& Informatics, University of Pennsylvania
}
\maketitle

\begin{abstract}
In this paper, we propose optimal transport rank regression for robust multivariate-response mean regression. The proposed method maps residual distributions to the uniform measure on the unit ball and constructs its loss from the resulting optimal transport ranks. This construction exploits cross-response dependence to improve efficiency while preserving population Fisher consistency and orthogonal invariance. Beyond these population properties, we establish statistical guarantees for the proposed method in both linear and nonparametric regression. For linear models, we establish asymptotic normality, characterise robustness, and develop orthogonally invariant asymptotic inference. We derive a distribution-uniform lower bound on asymptotic efficiency relative to least-squares and demonstrate efficiency gains over coordinatewise methods under several heavy-tailed and contaminated distributions. For nonparametric regression, we establish non-asymptotic \(L^2(P)\) convergence rates and robustness guarantees modulo additive constants using deep neural networks under possibly asymmetric heavy-tailed noise, attaining the minimax-optimal rate up to logarithmic factors. Numerical experiments demonstrate improvements in estimation accuracy and inferential efficiency over coordinatewise and radial robust methods, alongside stability under orthogonal transformations of the responses. An oral microbiome application further shows that the proposed method identifies more associations that are significant in all examined orthonormal response representations compared with each competing method.
\end{abstract}

\startlist{toc}
{
  \hypersetup{linkcolor=black}
  \printlist{toc}{}{}
}

\section{Introduction}\label{sec: introduction}

A common goal in genetic studies~\cite{park2024low,lit_chen2012}, neuroimaging~\cite{zhou2016regression}, and econometrics~\cite{lit_geweke1996} is to understand how covariates affect multiple scientifically related outcomes. Motivated by such applications, we develop estimation and inference procedures for robust multivariate-response mean regression. Compared with analysing each response separately, joint modelling can exploit cross-response dependence to improve estimation, prediction, and inference in linear models~\cite{lit_zellner1962,lit_breiman1997}. Similar gains are desirable when estimating complex, potentially nonlinear regression functions.

Deep neural networks (DNNs) provide flexible models for such relationships and can exploit low-dimensional structure to mitigate the curse of dimensionality in nonparametric regression~\cite{bauer2019deep,schmidt2020nonparametric,kohler2021rate}. Their expressive capacity has also enabled new statistical approaches accommodating multivariate responses, including engression~\cite{shen2025engression}, Wasserstein generative regression~\cite{song2026wasserstein}, multiple-output quantile regression~\cite{feldman2023calibrated}, and deep conditional sampling~\cite{zhou2023deep}. However, expressive capacity alone does not guarantee statistically optimal estimation under heavy-tailed noise. Heavy tails are pervasive in modern high-dimensional data~\cite{fan2021shrinkage} and can undermine conventional non-robust procedures, such as least-squares (LS) regression~\cite{sun2020adaptive}. Recent analysis of regression with deep rectified linear unit (ReLU) networks under heavy-tailed noise highlights a delicate interplay between approximation error, robustification bias, and statistical error~\cite{fan2024noise}. For multivariate-response regression, the challenge is therefore to retain the efficiency gains of joint modelling while translating the flexibility of DNNs into minimax-optimal estimation under potentially asymmetric heavy-tailed noise.

Alongside efficiency gains and minimax-optimal convergence guarantees, a robust procedure for multivariate-response mean regression must preserve the population target. Fisher consistency formalises this requirement: the true mean function should minimise the population risk~\cite{cox1974theoretical,hampel1986robust}. Without it, the minimiser of the population risk can differ from the parameter of interest, and this discrepancy can compromise both estimation and inference. A natural multivariate construction applies a robust loss, such as the Huber loss, to the Euclidean norm of the residual vector~\cite{she2017robust,lopuhaa2023sestimation}. Under asymmetric noise, however, the population risk of such a radial robust loss need not be minimised at the true mean function. Alternatively, Fisher-consistent univariate procedures, such as antitonic score matching (ASM)~\cite{feng2026optimal} and composite quantile regression (CQR)~\cite{zou2008composite}, can be applied separately to each response. CQR has also been extended to nonparametric regression through local polynomial smoothing~\cite{kai2010local}. These coordinatewise extensions can retain Fisher consistency but do not exploit cross-response dependence. Recently, \textcite{yang2024multiple} linked the CQR loss to optimal transport (OT) and used this link to develop multiple-output composite quantile regression (MCQR) for multivariate-response mean regression. Because its loss depends on the joint residual distribution, MCQR can exploit cross-response dependence to improve estimation accuracy. \textcite{yang2024multiple} established convergence bounds for linear models, which deteriorate with the response dimension and, under moment conditions alone, with the covariate dimension. To our knowledge, convergence guarantees for nonparametric extensions of MCQR, for example with DNNs, remain to be established.

Joint modelling should also yield statistical conclusions that do not depend on an arbitrary choice of response coordinates. Orthogonal invariance expresses this requirement~\cite{lehmann1959testing,rousseeuw2004robust}. For example, compositional responses admit different orthonormal representations chosen to facilitate scientific interpretation~\cite{egozcue2003isometric}, yet regression conclusions should remain independent of that choice~\cite{aitchison1986statistical}. Coordinatewise procedures generally violate this requirement, yielding different estimates and testing conclusions across response bases even after transformation back to the original coordinates. Such representation dependence has practical consequences, including changes in the sets of genes identified as significant in spatial transcriptomics~\cite{su2025stance} and in testing and model-selection results in phylogenetic studies~\cite{adams2018multivariate}. Existing multivariate rank regression methods achieve orthogonal invariance~\cite{visuri2003affine,ollila2003estimates} but can lose Fisher consistency under asymmetric noise and have unbounded influence under noise contamination. Their constructions also target linear models and do not readily extend to nonlinear classes such as DNNs.

To address these challenges, we propose optimal transport rank regression (OTRR), a robust framework for joint mean regression that accommodates both linear models and nonlinear candidate classes. For each candidate regression function, OTRR maps the joint residual distribution to the uniform measure on the unit ball and defines its loss through the resulting bounded OT ranks. This distribution-adaptive construction exploits cross-response dependence while preserving Fisher consistency. The bounded OT ranks underpin our robustness guarantees, and the spherical reference measure yields orthogonal invariance of the population risk. As in classical rank regression, OTRR identifies the regression function up to an additive constant. For linear models, we establish asymptotic inference and efficiency guarantees. For deep ReLU regression, we show that the $L^2(P)$ error, measured modulo additive constants, attains the minimax-optimal rate up to logarithmic factors, and we provide non-asymptotic robustness guarantees.

Our main contributions are summarised as follows.

\begin{enumerate}
    \item  \textbf{Fisher consistency and orthogonal invariance as intrinsic properties.}
    We establish that OTRR simultaneously achieves Fisher consistency and orthogonal invariance over general candidate classes. Specifically, Theorem~\ref{thm: orthogonal-equivariance} identifies the true mean function as a population risk minimiser and shows that the risk is invariant under orthogonal transformations of the response. These guarantees hold whenever the noise is independent of the covariates and absolutely continuous with finite variance; they require no symmetry or ellipticity assumptions and no linear model specification. In contrast, Theorem~\ref{thm:radial-impossibility} establishes limitations on achieving Fisher consistency with radial robust losses.

    \item \textbf{Robust estimation and efficiency guarantees for linear models.}
    We develop asymptotic theory for fixed-dimensional linear OTRR. Theorem~\ref{thm:full-linear-asymptotic-normality} establishes a $\sqrt{n}$-Gaussian limit with an orthogonally equivariant asymptotic covariance, yielding asymptotically basis-independent Wald tests for linear contrasts. Corollary~\ref{cor:full-bounded-response-asymptotic-influence} establishes bounded asymptotic influence conditional on the covariates, providing robustness to outlying responses. Our efficiency analysis shows that, uniformly over noise distributions, the asymptotic efficiency of OTRR relative to LS is at least $84\%$, with the lower bound approaching one as the response dimension increases. Comparisons with marginal maximum likelihood estimators further show that, by exploiting cross-response dependence, OTRR can be asymptotically more efficient than any regular coordinatewise method under several heavy-tailed and contaminated noise distributions.

    \item \textbf{Minimax-optimal nonparametric regression with deep ReLU networks.}
    We establish that deep ReLU OTRR with entropic OT-rank estimation attains, up to logarithmic factors, the minimax-optimal rate of convergence in $L^2(P)$ modulo additive constants over hierarchical composition classes, under potentially asymmetric heavy-tailed noise. Under our conditions on the noise density, the convergence exponent reflects the target function's low-dimensional compositional structure and contains no noise-moment parameter. We also derive high-probability error bounds with sub-Gaussian dependence on the confidence parameter, combining minimax optimality with non-asymptotic robustness.
\end{enumerate}

The numerical studies in Section~\ref{sec: Numerical Studies} demonstrate the practical benefits of OTRR.
In simulations, OTRR improves estimation accuracy and inferential efficiency relative to competing methods while remaining nearly orthogonally invariant in finite samples.
The National Health and Nutrition Examination Survey (NHANES) application and a complementary semi-synthetic experiment further illustrate the advantages of OTRR as an orthogonally invariant method that exploits cross-response dependence to support stable scientific discoveries and improve testing power.
Together, these results support OTRR as a theoretically grounded and practically viable approach to robust multivariate-response mean regression.

The remainder of the paper is organised as follows.
Section~\ref{sec: Method} introduces OTRR.
Section~\ref{sec: Fisher Consistency and Orthogonal Invariance} establishes its population properties and contrasts them with the limitations of radial losses.
Section~\ref{sec:full-linear-asymptotic-normality} develops the asymptotic theory for linear models.
Section~\ref{sec:relu-entropic} establishes convergence and robustness guarantees for nonparametric regression with deep ReLU networks.
Section~\ref{sec: Numerical Studies} presents simulations and the NHANES oral microbiome application.
Section~\ref{sec: Conclusion} concludes the paper.
The Supplementary Material~\cite{supplement} provides additional theoretical results and discussions, complete proofs, implementation details, and further numerical results.

Before proceeding, we collect the notations used throughout the paper. For $n\in\NN$, write $[n]:=\{1,\ldots,n\}$. We use $\inner{\cdot,\cdot}$ for the Euclidean inner product and $\|\cdot\|_1$, $\|\cdot\|_2$, $\|\cdot\|_{\F}$, and $\|\cdot\|_{\op}$ for the $\ell_1$, $\ell_2$, Frobenius, and operator norms, respectively. Let $B_d(0,1):=\{\ubf\in\RR^d:\|\ubf\|_2\le1\}$ be the unit ball. For $\mathbf v=(v_1,\ldots,v_n)\in(\RR^d)^n$, define $\|\mathbf v\|_{\ell^n}^2:=n^{-1}\sum_{i=1}^n\|v_i\|_2^2$. For a vector-valued function $f:\Xcal\to\RR^d$, define $\|f\|_{2,L^2(P)}^2:=\EE\|f(\xbf)\|_2^2$ and $\|f\|_{2,L^\infty}:=\sup_{\xbf\in\Xcal}\|f(\xbf)\|_2$. We write $S_\#P$ for the push-forward of $P$ under a map $S$. For a matrix $A$, $\operatorname{vec}(A)$ stacks its columns, and $\otimes$ denotes the Kronecker product. Finally, $\lesssim$ denotes an inequality up to a universal constant unless otherwise specified, and $\lesssim_{\log}$ additionally suppresses logarithmic factors.

\section{Optimal Transport Rank Regression}\label{sec: Method}

In this section, we introduce the OTRR method. Section~\ref{subsec:method-motivation} uses an inner-product representation of regression losses to examine Fisher consistency and orthogonal invariance within robust multivariate-response regression. Section~\ref{subsec:method-criterion} defines the population and empirical risks and the corresponding OTRR estimators. Some preliminaries on robust regression and multivariate-response regression are given in Section~\ref*{sec: Preliminaries and Related Literature} of the Supplementary Material.

\subsection{Motivation: Principles for Robust Multivariate Regression}\label{subsec:method-motivation}

We seek a robust multivariate-response regression procedure that preserves both orthogonal invariance and Fisher consistency. To understand the constraints imposed by these requirements, we examine an inner-product representation of regression losses.

We first introduce notation for the regression model. Consider i.i.d.\ observations $(\xbf_i,\ybf_i)\in\RR^p\times\RR^d$, \(i\in[n]\) satisfying, for some target function \(f_0:\RR^p\to\RR^d\):
\begin{equation}\label{eq: data generating process of OT_rank regression}
    \ybf_i = f_0(\xbf_i)+\xibf_i,\quad \text{for }\xibf_i\ind \xbf_i,\ \EE[\xibf_i]=0.
\end{equation}
Writing $(\xbf_i,\ybf_i)\sim P$, let $R=\ybf-f(\xbf)\in\RR^d$ denote the residual for a candidate function $f$. Many multivariate regression losses admit the inner-product representation
\(
    \ell(R):=\langle R,\delta(R)\rangle,
\) for $ R\in\RR^d$.
Here, $\delta:\RR^d\to\RR^d$ is a residual transform that may depend on \(f\) and \(P\). For instance, LS has $\ell^{\mathrm{LS}}(R)=\|R\|_2^2$ and $\delta^{\mathrm{LS}}(R)=R$. Coordinatewise Huber regression, with tuning parameter $\tau>0$, has loss $\ell_\tau(R):=\sum_{j=1}^d\ell_\tau(R_j) =\langle R,\delta_\tau^{\mathrm H}(R)\rangle$, where $\ell_\tau(u)=u^2$ for $|u|\le\tau$ and $\ell_\tau(u)=2\tau|u|-\tau^2$ otherwise.
The corresponding residual transform satisfies $(\delta_\tau^{\mathrm H}(R))_j=R_j$ for $|R_j|\le\tau$ and $(\delta_\tau^{\mathrm H}(R))_j=2\tau\,\mathrm{sgn}(R_j)-\tau^2/R_j$ otherwise.

As Lipschitz losses provide a useful route to non-asymptotic robustness under heavy-tailed noise, we consider uniformly bounded residual transforms, ensuring that $\ell(R)=\langle R,\delta(R)\rangle$ grows at most linearly in $\|R\|_2$. The coordinatewise Huber transform $\delta_\tau^{\mathrm H}$ satisfies this boundedness requirement, whereas the LS transform $\delta^{\mathrm{LS}}(R)=R$ does not.

Within this robust regression framework, we additionally require Fisher consistency and orthogonal invariance.
\begin{enumerate}
    \item \textbf{Fisher consistency.} Robustification should limit sensitivity to extreme observations without changing the population target. Fisher consistency formalises this requirement by ensuring that the true mean function minimises the population risk~\cite{cox1974theoretical,hampel1986robust}. For a fixed residual transform $\delta$ that does not depend on $f$ or $P$, the first-order condition at the true mean function involves  $\EE\bigl[\delta(\xibf)+\{\nabla\delta(\xibf)\}^\top\xibf\bigr]=0$. Thus, apart from special cases such as LS, a fixed residual transform generally fails Fisher consistency for all mean-zero noise distributions. Theorem~\ref{thm:radial-impossibility} in Section~\ref{sec: Fisher Consistency and Orthogonal Invariance} further formalises the limitations of radial losses, motivating a distribution-dependent and generally non-radial loss. Failure of Fisher consistency can introduce population bias, thereby compromising the validity of hypothesis tests~\cite{chen2020robust}.

    \item \textbf{Orthogonal invariance.}
    In multivariate-response regression, changing the orthonormal response basis should change only the representation of the regression problem, not its statistical conclusions~\cite{rousseeuw2004robust,visuri2003affine,ollila2003estimates}. Orthogonal invariance formalises this requirement at the loss level through $\ell(QR)=\ell(R)$ for every orthogonal matrix $Q\in\RR^{d\times d}$. For a fixed residual transform, a sufficient condition is that $\delta$ commutes with orthogonal transformations:
    \begin{equation}\label{eq: sufficient condition for orthogonal invariance}
        \delta(QR)=Q\delta(R)\qquad\text{for every }R\in\RR^d\text{ and every orthogonal }Q\in\RR^{d\times d}.
    \end{equation}
    LS satisfies this condition, whereas coordinatewise Huber regression does not because its truncation depends on the chosen coordinate axes.

    Orthogonal invariance ensures equivariant estimation: fitting after a response rotation and mapping back yields the original estimate. It also has direct implications for inference: without it, changing coordinates can alter statistical conclusions about the same scientific question~\cite{adams2018multivariate,su2025rotation}. Coordinatewise robust regression methods can also miss multivariate outliers~\cite{rousseeuw2004robust}. Our oral-microbiome analysis in Section~\ref{sec: Real Data Application} further illustrates representation-dependent discoveries for these methods, while the accompanying semi-synthetic experiment provides complementary evidence on testing power.
\end{enumerate}
These considerations motivate a distribution-adaptive and non-radial approach to robust regression that preserves orthogonal invariance without sacrificing Fisher consistency.

\subsection{OT-Rank Regression: Risks and Estimators}\label{subsec:method-criterion}\label{subsec:optimal-transport}

The preceding discussion motivates a uniformly bounded residual transform that adapts to the residual distribution while respecting orthogonal changes of response coordinates. OT provides a natural construction by transporting each candidate residual distribution to a bounded, spherical reference distribution.

To formalise this construction, let $\mu$ and $\nu$ be probability measures on $\RR^d$ with finite second moments. An OT map from $\mu$ to $\nu$ minimises the transportation cost $\int\|x-T(x)\|_2^2\,\mu(\dd x)$ over measurable maps satisfying $T_\#\mu=\nu$. When $\mu$ is absolutely continuous, Brenier's theorem guarantees that this map uniquely exists and is the gradient of a convex function~\cite{brenier1987decomposition}. This convex-gradient characterisation extends the monotonicity underlying univariate ranks: in one dimension, the non-decreasing distribution function of an absolutely continuous law is its OT map to $\operatorname{Unif}(0,1)$. Analogously, a multivariate OT rank is defined by transporting $\mu$ to a fixed reference measure $\nu$, such as the uniform distribution on the unit ball~\cite{chernozhukov2017monge,hallin2021distribution,ghosal2022multivariate}. Coordinatewise ranks underlie semiparametric multi-response methods such as multitask quantile regression under the transnormal model~\cite{fan2016multitask}, while OT ranks instead transform the vector jointly as in the cyclically monotone copula Gaussian graphical model~\cite{zhang2026copula}.

OT ranks thus provide a distribution-adaptive residual transform through the choice of a suitable reference distribution. Unless stated otherwise, the OTRR reference distribution is the normalised Lebesgue measure on $B_d(0,1)$. Let $T_f:\RR^d\to\RR^d$ transport the law of $\ybf-f(\xbf)$ to this reference distribution. We define the population risk of OTRR by
\begin{equation}\label{eq: population loss function of OT-rank regression}
    \Lcal(f):=\EE[(\ybf-f(\xbf))^\top T_f(\ybf-f(\xbf))].
\end{equation}
This OTRR loss has the structural features anticipated in Section~\ref{subsec:method-motivation}. The OT map $T_f$ is distribution-adaptive and generally non-radial. Since the reference distribution is supported on the unit ball, $\|T_f(\cdot)\|_2\le1$, yielding the bounded residual transform motivated in Section~\ref{subsec:method-motivation}. Finally, the spherical reference distribution ensures that the OT-rank residual transforms satisfy the distribution-adaptive analogue of~\eqref{eq: sufficient condition for orthogonal invariance}, yielding orthogonal invariance of the OTRR population risk.

As in classical rank regression, OTRR identifies the regression function up to an additive constant. Indeed, for any constant vector $\abf\in\RR^d$ and $f_{\abf}:=f+\abf$, the residual law of $\ybf-f_{\abf}(\xbf)$ is a translation of the residual law of $\ybf-f(\xbf)$. Hence, the corresponding OT map satisfies $T_{f_{\abf}}(\ubf)=T_f(\ubf+\abf)$ almost everywhere, and $\Lcal(f_{\abf}) = \EE\big[(\ybf-f(\xbf)-\abf)^\top T_f(\ybf-f(\xbf))\big] = \Lcal(f)$.

For each candidate $f$, let $\widehat T_f$ estimate $T_f$ from the residuals $\{\ybf_i-f(\xbf_i)\}_{i=1}^n$. Our formulation does not prescribe a particular construction of $\widehat T_f$ and accommodates plug-in~\cite{ghosal2022multivariate,deb2021rates,manole2021plugin}, dual~\cite{hutter2021minimax,gunsilius2022convergence,divol2025optimal}, and entropic estimators~\cite{pooladian2021entropic,goldfeld2024limit}. The empirical risk and estimator of OTRR are then defined by
\begin{equation}\label{eq:full-empirical-risk}
    \Lcal_n(f) := \frac1n\sum_{i=1}^n(\ybf_i-f(\xbf_i))^\top\widehat T_f\{\ybf_i-f(\xbf_i)\}, \qquad \widehat f_n\in\argmin_{f\in\Fcal_n}\Lcal_n(f).
\end{equation}

This formulation accommodates general candidate classes. In this work, we consider two concrete examples for theoretical analysis: linear models and deep ReLU networks.

\noindent\textbf{Linear models.} We specialise model~\eqref{eq: data generating process of OT_rank regression} to
\begin{equation}\label{eq: data generating process of OT_rank regression for linear model}
    \ybf_i=\Bbf_0\xbf_i+\xibf_i, \qquad \Bbf_0\in\RR^{d\times p}, \quad \xibf_i\ind\xbf_i, \quad \EE[\xbf_i]=\EE[\xibf_i]=0.
\end{equation}
Here, $\EE[\xbf_i]=0$ resolves the additive-constant non-identifiability issue discussed above.

For a candidate coefficient matrix $\Bbf$, let $T_{\Bbf}$ transport the law of $\ybf-\Bbf\xbf$ to the uniform distribution on $B_d(0,1)$, write $T_0:=T_{\Bbf_0}$, and let $\widehat T_{\Bbf}$ estimate $T_{\Bbf}$ from $\{\ybf_i-\Bbf\xbf_i\}_{i=1}^n$. The population and empirical risks in~\eqref{eq: population loss function of OT-rank regression} and~\eqref{eq:full-empirical-risk} become $\Lcal(\Bbf) := \EE\bigl[(\ybf_1-\Bbf\xbf_1)^\top T_{\Bbf}(\ybf_1-\Bbf\xbf_1)\bigr]$ and $\Lcal_n(\Bbf) := \frac1n\sum_{i=1}^n(\ybf_i-\Bbf\xbf_i)^\top\widehat T_{\Bbf}(\ybf_i-\Bbf\xbf_i)$, respectively.

Let $\widetilde\Bbf_n$ be a pilot estimator of $\Bbf_0$; one possible construction applies a classical regression method separately to each response coordinate. Suppose that $\widetilde\Bbf_n$ is consistent in the sense that $\|\widetilde\Bbf_n-\Bbf_0\|_\F=o_\PP(r_n)$ for some $r_n\downarrow0$. We define the linear OTRR estimator by
\begin{equation}\label{eq:localized-linear-ot-rank-estimator}
    \widehat\Bbf_n \in \argmin_{\Bbf:\|\Bbf-\widetilde\Bbf_n\|_\F\le r_n} \Lcal_n(\Bbf).
\end{equation}
Thus, the OTRR estimator $\widehat\Bbf_n$ locally refines the pilot $\widetilde\Bbf_n$.

\noindent\textbf{Deep ReLU networks.} We consider the class of  deep ReLU networks with width $W$ and depth $D$ with uniformly bounded outputs:
\begin{equation}\label{eq:relu-entropic-class}
    \Fcal_n:=\Bigl\{\operatorname{clip}_{L_1}\circ f:\, f=\mathsf{Aff}_{D+1}\circ\sigma\circ\mathsf{Aff}_{D}\circ\sigma\circ\cdots\circ\mathsf{Aff}_2\circ\sigma\circ\mathsf{Aff}_1(x)\Bigr\},
\end{equation}
where $\sigma(x)=0\lor x$ is applied entrywise, $\mathsf{Aff}_i(x)=A_i x+b_i$ with $A_i\in\RR^{d_i\times d_{i-1}}$ and $b_i\in\RR^{d_i}$, and $(d_0,d_1,\ldots,d_D,d_{D+1})=(p,W,\ldots,W,d)$. 
For $L_1>0$, let $\operatorname{clip}_{L_1}$ act coordinatewise by $u\mapsto(-L_1)\lor(u\land L_1)$. Then every $f\in\Fcal_n$ satisfies $\|f\|_{2,L^\infty([0,1]^p)}\le\sqrt{d}\,L_1$. Section~\ref{sec:relu-entropic} defines the deep ReLU OTRR estimator over $\Fcal_n$ using entropic OT ranks.

\begin{remark}[Connection and comparison with MCQR]\label{rem:method-comparison}
OTRR and MCQR~\cite{yang2024multiple} share the philosophy of using OT to exploit cross-response dependence in multivariate-response regression. The univariate methods they generalize, rank regression and CQR respectively, have equivalent population risks, and this equivalence extends to OTRR and MCQR under a common reference distribution. Section~\ref*{sec:supp-rank-mcqr-comparison} of the Supplementary Material gives the details.

Despite this population-level equivalence, the two approaches lead to different empirical constructions and theoretical analyses. OTRR plugs estimated OT ranks into the inner-product form of its risk and does not prescribe a particular rank estimator. This explicit form lets us analyze the resulting estimators through regularity conditions on the estimated ranks. By contrast, MCQR defines its empirical risk as the Wasserstein product of the empirical residual distribution and an empirical reference distribution, and its existing analysis relies on uniform control of empirical Wasserstein distances. Under their respective assumptions, our analysis yields a $\sqrt{n}$-Gaussian limit for linear OTRR, whereas the available MCQR theory provides convergence bounds whose rates deteriorate with dimension.

The freedom to choose the OT-rank estimator also gives computational flexibility. With entropic OT ranks, for example, OTRR can be fitted by gradient-based optimization, including for nonlinear classes such as DNNs, at a cost per step proportional to the product of the sample size and the number of reference points. The MCQR procedure of \textcite{yang2024multiple}, by comparison, solves a linear program with that many variables, which can impose a substantial computational burden for large samples.
\end{remark}

\section{Fisher Consistency and Orthogonal Invariance}\label{sec: Theory}\label{sec: Fisher Consistency and Orthogonal Invariance}
This section first establishes Fisher consistency and orthogonal invariance of OTRR, then contrasts these properties with the limitations of radial losses. Unless stated otherwise, the reference random vector $\zbf$ is uniformly distributed on $B_d(0,1)$.

We begin with the OTRR population risk. Fisher consistency means that $f_0$ minimises this risk under model~\eqref{eq: data generating process of OT_rank regression}. To formalise orthogonal invariance, for an orthogonal matrix $Q\in\RR^{d\times d}$, let $T_h^{(Q)}$ denote the OT map from the law of $Q\ybf-h(\xbf)$ to that of $\zbf$, and define
\[
    \Lcal_Q(h):=\EE\Big[(Q\ybf-h(\xbf))^\top T_h^{(Q)}(Q\ybf-h(\xbf))\Big].
\]
Orthogonal invariance then requires that $\Lcal_Q(Qf)=\Lcal(f)$ for all $f$. In other words, the OTRR should not depend on the orthonormal coordinate system of the response.

To show that these properties are intrinsic to OTRR, we impose as few regularity assumptions as possible. In particular, we assume only absolutely continuous noise with finite second moments in model~\eqref{eq: data generating process of OT_rank regression}, allowing the use of $2$-Wasserstein distance.
\begin{theorem}\label{thm: orthogonal-equivariance}
    Assume that $\zbf$ has a spherical distribution with finite variance. Suppose, for the model~\eqref{eq: data generating process of OT_rank regression}, that $\EE[\|\xibf\|_2^2]<\infty$, $\xibf$ is absolutely continuous, and $f_0\in L^2(P;\RR^d)$. Then,
    \[
        \Lcal_Q(Qf)=\Lcal(f)\geq \Lcal(f_0),\quad\text{ for any }f\in L^2(P;\RR^d)\text{, and any orthogonal }Q\in\RR^{d\times d}.
    \]
\end{theorem}
The inequality $\Lcal(f)\geq\Lcal(f_0)$ establishes Fisher consistency: the OTRR population risk is minimised at the target function $f_0$. The identity $\Lcal_Q(Qf)=\Lcal(f)$ establishes orthogonal invariance of this risk, so its minimisers transform equivariantly: if $f_0$ minimises $\Lcal$, then $Qf_0$ minimises $\Lcal_Q$. Here, Fisher consistency does not require \(f_0\) to be the unique population risk minimiser; as shown in Section~\ref{sec: Method}, the risk is invariant under additive shifts of the regression function. Although this orthogonal invariance identity need not hold exactly for the empirical risk, the numerical studies in Section~\ref{sec: Numerical Studies} show that the OTRR estimator remains nearly orthogonally invariant in finite samples. For linear models, Section~\ref{sec:full-linear-asymptotic-normality} establishes equivariance of the estimator's asymptotic distribution and covariance, yielding asymptotically invariant hypothesis tests and power across orthonormal response representations.

We next formalise the limitations of radial losses discussed in Section~\ref{subsec:method-motivation}.
\begin{theorem}\label{thm:radial-impossibility}
    Let $\phi:[0,\infty)\to\RR$ be a radial loss that is continuous on $[0,\infty)$ and twice continuously differentiable on $(0,\infty)$. If the loss is allowed to depend on the law of $\xibf$, we write it as $\phi_\xibf$. For $\thetabf\in\RR^d$, define the corresponding population risks by $L(\thetabf):=\EE\big[\phi(\|\xibf-\thetabf\|_2)\big]$ and $L_\xibf(\thetabf):=\EE[\phi_\xibf(\|\xibf-\thetabf\|_2)]$, respectively. Then,
    \begin{enumerate}
        \item if $\thetabf=0$ minimises $\thetabf\mapsto L(\thetabf)$ for every mean-zero law of $\xibf$, then there exist constants $c\ge0$ and $C\in\RR$ such that $\phi(r)=c\, r^2+C$ for all $r\ge0$.

        \item Suppose $d\ge2$. If $\thetabf=0$ minimises $\thetabf\mapsto L_\xibf(\thetabf)$ for every mean-zero law of $\xibf$ with $\EE\|\xibf\|_2^2=1$ and the family $\{\phi_{\xibf}\}$ is uniformly Lipschitz over these laws, then $\inf\lambda_{\min}\{\nabla^2L_{\xibf}(0)\}=0$, where the infimum is taken over all such laws for which $L_{\xibf}$ is twice differentiable at $0$.
    \end{enumerate}
\end{theorem}

Theorem~\ref{thm:radial-impossibility} clarifies the cost of achieving orthogonal invariance through radial robust losses. For fixed radial losses, Fisher consistency permits only least-squares or degenerate constant losses. Allowing distribution dependence may recover Fisher consistency, but a uniformly Lipschitz family cannot maintain a positive curvature lower bound uniformly over mean-zero noise distributions, even with their variance fixed at one. 

Together, these guarantees distinguish OTRR from coordinatewise robust losses that generally lack orthogonal invariance and radial robust losses that may fail Fisher consistency.

\section{Theoretical Analysis for Linear Models}\label{sec:full-linear-asymptotic-normality}

We now establish asymptotic normality and efficiency comparisons for the linear OTRR estimator~\eqref{eq:localized-linear-ot-rank-estimator} under model~\eqref{eq: data generating process of OT_rank regression for linear model}. Unlike LS and Huber losses, which depend on $\Bbf$ only through the residuals, the OTRR loss depends on $\Bbf$ additionally through the candidate-dependent OT-rank estimator $\widehat T_{\Bbf}$. Establishing asymptotic normality therefore requires consistency of $\widehat T_{\Bbf}$ and stability under perturbations of its source samples. The following assumption formalises these requirements: consistency and stability of the OT-rank estimator control the first-order behaviour of the empirical OTRR risk, while the remaining conditions govern its second-order behaviour near $\Bbf_0$.

\begin{assumption}\label{asum:full-linear-asymptotic-normality}
    Under the preceding setup, assume $\Sigma_\xbf:=\EE[\xbf_i\xbf_i^\top]\succ0$ and the pilot estimator satisfies $\|\widetilde\Bbf_n-\Bbf_0\|_\F=o_\PP(r_n)$ for a sequence $r_n\downarrow0$ with $\sqrt n\,r_n\to\infty$. In addition,
    \begin{enumerate}[label=(\roman*),ref=\roman*]
        \item\label{item:full-map-l2-convergence} \textbf{Consistency of OT-rank estimator.} Assume $\widehat T_0:=\widehat T_{\Bbf_0}$ consistently estimates $T_0$ in the sense that $n^{-1}\sum_{i=1}^n\|\widehat T_0(\xibf_i)-T_0(\xibf_i)\|_2^2=o_\PP(1)$.
        \item \textbf{Stability of OT-rank estimator.}
        For $\mathbf v=(v_1,\ldots,v_n)\in (\RR^d)^n$, let $\widehat T_{t,\mathbf v}$ denote the OT-rank estimator constructed from the perturbed residuals $(\xibf_1+tv_1,\ldots,\xibf_n+tv_n)$. Assume
        \begin{equation}\label{eq:full-score-l2-consistency}
        \begin{aligned}
            \sup_{\|\mathbf v\|_{\ell^n}\le1}\left| \frac1n\sum_{i=1}^n \xibf_i^\top \left. \frac{\dd}{\dd t} \widehat T_{t,\mathbf v}(\xibf_i+tv_i)\right|_{t=0}\right|=o_\PP(1).
        \end{aligned}
        \end{equation}

        \item \textbf{Averaged local curvature.} Suppose $\EE\|\nabla T_0(\xibf)\|_{\op}<\infty$ and $\EE[\nabla T_0(\xibf)]\succ0$. Set $H_0:=\Sigma_\xbf\otimes\EE\{\nabla T_0(\xibf)\}$. For every fixed $V\in\RR^{d\times p}$, with $h:=\operatorname{vec}(V)$, assume
        \begin{equation}\label{eq:full-linear-pointwise-hessian}
            2\int_0^1(1-t)\,h^\top\nabla^2_{\operatorname{vec}(\Bbf)}\Lcal_n\left(\Bbf_0+\frac{tV}{\sqrt n}\right)h\,\dd t\xrightarrow{\PP}h^\top H_0h.
        \end{equation}
        \item\label{item:full-linear-convex-approximation} \textbf{Local convex approximation.} Write $\Dcal_n:=\{\Bbf:\|\Bbf-\Bbf_0\|_\F\le2r_n\}$. There exists a random convex function $G_n:\Dcal_n\to\RR$ such that $\sup_{\Bbf\in\Dcal_n}|\Lcal_n(\Bbf)-G_n(\Bbf)|=o_\PP(n^{-1})$.
    \end{enumerate}
\end{assumption}

Assumption~\ref{asum:full-linear-asymptotic-normality} accommodates a broad class of OT-rank estimators without prescribing a particular construction of $\widehat T_{\Bbf}$. Conditions~(\ref{item:full-map-l2-convergence}) and~\eqref{eq:full-score-l2-consistency} provide zeroth- and first-order control, respectively: the former requires consistency of the estimated OT ranks at the residuals, while the latter controls their pathwise derivatives under residual perturbations. Conditions~\eqref{eq:full-linear-pointwise-hessian} and~(\ref{item:full-linear-convex-approximation}) impose averaged local curvature and local convex approximation, respectively. Both may follow from uniform Hessian convergence of empirical risk to $H_0\succ0$ on $\Dcal_n$. We further verify these conditions for entropic OT estimators, the primary approach used in our numerical studies, as detailed in Section~\ref*{sec: Verification for Entropic OT Estimators} of the Supplementary Material.

Condition~\eqref{eq:full-score-l2-consistency} is central to the score convergence underlying Theorem~\ref{thm:full-linear-asymptotic-normality}. It makes the additional derivative term arising from the dependence of OT-rank estimators on the candidate residual distribution asymptotically negligible. Importantly, this condition is motivated by an exact population identity for OT ranks, rather than imposed solely for technical convenience. For a sufficiently regular velocity field $v$, let $P_t=(\operatorname{id}+tv)_\#P_{\xibf}$ be the distribution of $\xibf+tv(\xibf)$, and let $T_t$ transport $P_t$ to the reference distribution. Under the $L^2$ pathwise differentiability condition in Proposition~\ref*{prop:population-ot-score-cancellation} of the Supplementary Material,
\begin{equation}\label{eq:main-population-ot-score-cancellation}
    \EE\!\left[\xibf^\top\left.\frac{\dd}{\dd t}T_t\{\xibf+tv(\xibf)\}\right|_{t=0}\right]=0.
\end{equation}
This identity is an infinitesimal multivariate analogue of the invariance of univariate ranks under increasing transformations. In one dimension, if $h_t(x):=x+tv(x)$ is strictly increasing and $F_t$ is the distribution function of $h_t(\xibf)$, then $F_t\{h_t(\xibf)\}=F_0(\xibf)$ almost surely: ranks remain unchanged along each strictly increasing path. In the multivariate setting, identity~\eqref{eq:main-population-ot-score-cancellation} gives the corresponding infinitesimal analogue after weighting by $\xibf$ and taking expectation. Its derivation is provided in Section~\ref*{sec:population-ot-score-cancellation} of the Supplementary Material.

Based on these conditions, the following theorem establishes asymptotic normality of the linear OTRR estimator, which forms the basis for the subsequent efficiency comparisons.

\begin{theorem}\label{thm:full-linear-asymptotic-normality}
    Suppose that Assumption~\ref{asum:full-linear-asymptotic-normality} holds, and define $H_0:=\Sigma_\xbf\otimes\EE\{\nabla T_0(\xibf)\}$ and $\Omega_0:=\Sigma_\xbf\otimes\EE\{T_0(\xibf)T_0(\xibf)^\top\}$. Then, the estimator $\widehat\Bbf_n$ defined in~\eqref{eq:localized-linear-ot-rank-estimator} satisfies
    \[
        \sqrt n\,\operatorname{vec}(\widehat\Bbf_n-\Bbf_0)\rightsquigarrow N(0,H_0^{-1}\Omega_0H_0^{-1}).
    \]
\end{theorem}

Theorem~\ref{thm:full-linear-asymptotic-normality} establishes $\sqrt{n}$-asymptotic normality for linear OTRR. The following corollary shows that the resulting asymptotic influence function is bounded in the residuals for each fixed covariate value, providing a robustness guarantee. Therefore, residuals cannot have arbitrarily large first-order effects on linear OTRR.

\begin{corollary}\label{cor:full-bounded-response-asymptotic-influence}
    Under Assumption~\ref{asum:full-linear-asymptotic-normality}, let $\operatorname{IF}_0(\xbf,\xibf)$ denote the asymptotic influence function of the estimator $\widehat\Bbf_n$ in~\eqref{eq:localized-linear-ot-rank-estimator}. Then, $\sup_{\xibf}\|\operatorname{IF}_0(\xbf,\xibf)\|_2=\bigl\|\EE[\nabla T_0(\xibf)]^{-1}\bigr\|_{\op}\|\Sigma_\xbf^{-1}\xbf\|_2$.
\end{corollary}

We then develop Wald-type inference results. First, we show that the Gaussian limit in Theorem~\ref{thm:full-linear-asymptotic-normality} respects orthogonal changes of the response basis. Write $V_0:=H_0^{-1}\Omega_0H_0^{-1}$. For an orthogonal matrix $Q$, consider OTRR with response $Q\ybf$ and let $G_Q:=I_p\otimes Q$ and $\Bbf_0^{(Q)}:=Q\Bbf_0$, so that $\operatorname{vec}(\Bbf_0^{(Q)})=G_Q\operatorname{vec}(\Bbf_0)$. The spherical reference distribution implies that the population Hessian, score covariance, and limiting covariance satisfy $H_0^{(Q)}=G_QH_0G_Q^\top$, $\Omega_0^{(Q)}=G_Q\Omega_0G_Q^\top$, and $V_0^{(Q)}=(H_0^{(Q)})^{-1}\Omega_0^{(Q)}(H_0^{(Q)})^{-1}=G_QV_0G_Q^\top$. Thus, the limiting covariance transforms equivariantly and the corresponding tests retain the same asymptotic size and power. Orthogonal invariance therefore protects not only coefficient estimation but also the resulting hypothesis-testing conclusions. The oral-microbiome analysis in Section~\ref{sec: Real Data Application} illustrates basis-dependent testing conclusions for methods lacking this property; the accompanying semi-synthetic experiment further compares their testing power.

Asymptotic confidence intervals for linear contrasts of $\Bbf_0$ can be constructed by estimating the asymptotic covariance $V_0$. Let $\widehat\xibf_i:=\ybf_i-\widehat\Bbf_n\xbf_i$, $i=1,\ldots,n$, denote the estimated residuals. A natural plug-in estimator of $H_0$ is $\widehat H_n:=\widehat\Sigma_\xbf\otimes\widehat J_n$, where $\widehat\Sigma_\xbf:=n^{-1}\sum_{i=1}^n\xbf_i\xbf_i^\top$ and $\widehat J_n:=n^{-1}\sum_{i=1}^n\nabla\widehat T_{\widehat\Bbf_n}(\widehat\xibf_i)$. To estimate $\Omega_0$, define $\widehat\psi_i:=\operatorname{vec}\{\widehat T_{\widehat\Bbf_n}(\widehat\xibf_i)\xbf_i^\top\}$ and let $\widehat\Omega_n:=n^{-1}\sum_{i=1}^n(\widehat\psi_i-\overline\psi)(\widehat\psi_i-\overline\psi)^\top$, where $\overline\psi:=n^{-1}\sum_{i=1}^n\widehat\psi_i$. The resulting plug-in estimator of the limiting covariance matrix is
\(
    \widehat V_n := \widehat H_n^{-1} \widehat\Omega_n \widehat H_n^{-\top}.
\)
For a nonzero contrast vector $\abf\in\RR^{dp}$, the corresponding Wald interval for $\abf^\top\operatorname{vec}(\Bbf_0)$ is
\begin{equation}\label{eq:full-wald-contrast-interval}
    \abf^\top\operatorname{vec}(\widehat\Bbf_n) \ \pm\ z_{1-\alpha/2} \sqrt{\frac{\abf^\top\widehat V_n\abf}{n}}.
\end{equation}
For $A\in\RR^{r\times dp}$ with $AV_0A^\top$ nonsingular, the corresponding confidence set for $A\operatorname{vec}(\Bbf_0)$ is
\begin{equation}\label{eq:full-wald-contrast-region}
    \left\{ \ubf\in\RR^r: n\bigl\{\ubf - A\operatorname{vec}(\widehat\Bbf_n)\bigr\}^\top (A\widehat V_nA^\top)^{-1} \bigl\{\ubf - A\operatorname{vec}(\widehat\Bbf_n)\bigr\} \leq \chi^2_{r,1-\alpha} \right\}.
\end{equation}

The following result establishes the asymptotic validity of these confidence sets under additional conditions ensuring consistency of the plug-in covariance estimator, which we verify for the entropic OT estimator in Section~\ref*{sec: Verification for Entropic OT Estimators} of the Supplementary Material.

\begin{corollary}\label{cor:full-wald-validity}
    Under Assumption~\ref{asum:full-linear-asymptotic-normality}, suppose additionally that $\widehat J_n\xrightarrow{\PP}\EE\{\nabla T_0(\xibf)\}$ and $\frac1n\sum_{i=1}^n\|\xbf_i\|_2^2 \bigl\|\widehat T_{\widehat\Bbf_n}(\widehat\xibf_i)-T_0(\xibf_i)\bigr\|_2^2 =o_\PP(1)$. Then $\widehat V_n\xrightarrow{\PP}V_0$. For every $\alpha\in(0,1)$, $\abf\ne0$ and matrix $A$ as above, \eqref{eq:full-wald-contrast-interval} and ~\eqref{eq:full-wald-contrast-region} have coverage probabilities converging to $1-\alpha$.
\end{corollary}

The limiting covariance in Theorem~\ref{thm:full-linear-asymptotic-normality} also permits comparisons of the asymptotic relative efficiency (ARE) of OTRR with classical estimators. Corollary~\ref{cor:full-are-lower-bound} first provides a distribution-uniform lower bound on its ARE relative to LS. In particular, OTRR uniformly retains at least approximately $84\%$ efficiency. Analogous in spirit to the efficiency guarantee for CQR~\cite{zou2008composite}, this result rules out an arbitrarily large efficiency loss relative to LS. Moreover, the lower bound converges to one as $d\to\infty$, indicating that the worst-case efficiency loss relative to LS vanishes with the response dimension.

\begin{corollary}\label{cor:full-are-lower-bound}
    Under the conditions of Theorem~\ref{thm:full-linear-asymptotic-normality}, suppose additionally that the distribution of $\xibf$ is absolutely continuous and has a positive-definite covariance matrix. Let $V_{\rm LS}$ and $V_{\rm OT}$ denote the asymptotic covariance matrices of LS and OTRR, respectively. Define the ARE by \( \mathrm{ARE}^{\rm OT}_{\rm LS}(d):=\{\det(V_{\rm LS}) / \det(V_{\rm OT})\}^{1/(dp)}.
    \) Then,
    \begin{equation}\label{eq:full-determinant-are-lower-bound}
        \mathrm{ARE}^{\rm OT}_{\rm LS}(d) \ge \underline c_d := \frac{4(d+2)(d+1)^2}{(3d+2)^3}\left\{\frac{\Gamma(3d/2+1)}{\Gamma(d+1)\Gamma(d/2+1)}\right\}^{2/d}\ge 0.8427.
    \end{equation}
    In particular, $\underline c_1=0.864$ and $\underline c_d\to1$ as $d\to\infty$.
\end{corollary}

Beyond the distribution-uniform ARE lower bound relative to LS, radial noise distributions permit explicit efficiency comparisons. Table~\ref{tab:full-AREs-for-radial-laws} reports OTRR's ARE relative to LS, the marginal maximum likelihood estimator (mMLE), and the correctly specified joint maximum likelihood estimator (MLE). The mMLE fits the true marginal likelihood separately to each response coordinate, providing the efficiency benchmark for regular coordinatewise methods; the joint MLE uses the true joint likelihood and provides the parametric benchmark. In each comparison, an ARE above one favours OTRR.

\begin{table}[!htb]
    \centering\setlength{\tabcolsep}{4pt}\renewcommand{\arraystretch}{1.2}
    \caption{AREs of OTRR for several radial noise distributions.}
    \label{tab:full-AREs-for-radial-laws}
    \scalebox{0.85}{
    \begin{tabular}{cccccccccc}
        \toprule
        & \multicolumn{3}{c}{$\mathrm{ARE}^{\rm OT}_{\rm LS}$}
        & \multicolumn{3}{c}{$\mathrm{ARE}^{\rm OT}_{\rm mMLE}$}
        & \multicolumn{3}{c}{$\mathrm{ARE}^{\rm OT}_{\rm MLE}$}\\ \midrule
        Radial law
        & $d=1$ & $d=10$ & $d=20$
        & $d=1$ & $d=10$ & $d=20$
        & $d=1$ & $d=10$ & $d=20$\\ \midrule
        Cauchy & $\infty$ & $\infty$ & $\infty$ & 0.608 & 1.076 & 1.163 & 0.608 & 0.636 & 0.637\\
        $t_2$ & $\infty$ & $\infty$ & $\infty$ & 0.867 & 1.163 & 1.225 & 0.867 & 0.814 & 0.802\\
        $t_3$ & 1.900 & 2.312 & 2.409 & 0.950 & 1.156 & 1.205 & 0.950 & 0.889 & 0.873\\
        $t_{10}$ & 1.054 & 1.127 & 1.150 & 0.996 & 1.066 & 1.087 & 0.996 & 0.992 & 0.981\\
        $N(0,I_d)$ & 0.955 & 0.979 & 0.986 & 0.955 & 0.979 & 0.986 & 0.955 & 0.979 & 0.986\\
        $\mathrm{CN}(0.05,25)$ & 1.814 & 1.968 & 1.992 & 0.931 & 1.010 & 1.022 & 0.931 & 0.940 & 0.951\\
        $\mathrm{CN}(0.10,25)$ & 2.412 & 2.769 & 2.815 & 0.889 & 1.020 & 1.037 & 0.889 & 0.901 & 0.916\\
        \bottomrule
    \end{tabular}}
    \par\smallskip
{\footnotesize\textit{Note:}\
$\mathrm{CN}(\pi,\tau^2):=(1-\pi)N(0,I_d)+\pi N(0,\tau^2I_d)$.\par}
\end{table}

The ARE relative to the mMLE generally increases with $d$ and exceeds one for many heavy-tailed and contaminated distributions, implying that OTRR outperforms any regular coordinatewise estimator in these settings. Its ARE relative to the joint MLE remains below one but can increase with $d$, particularly under contaminated noise settings. These comparisons demonstrate OTRR's efficiency gains from exploiting cross-response dependence.

\section{Theoretical Analysis for Deep ReLU Networks}\label{sec:relu-entropic}

Section~\ref{sec:full-linear-asymptotic-normality} establishes asymptotic normality for linear OTRR. We now derive non-asymptotic convergence rates and robustness guarantees for the nonparametric estimator~\eqref{eq:full-empirical-risk} over the deep ReLU network class~\eqref{eq:relu-entropic-class} under model~\eqref{eq: data generating process of OT_rank regression}. Here, non-asymptotic robustness means that the estimation error in $L^2(P)$ modulo additive constants admits a high-probability bound with sub-Gaussian dependence on the confidence parameter, even under heavy-tailed noise.

A key advantage of deep ReLU networks in nonparametric regression is their ability to exploit low-dimensional structures of target functions~\cite{bauer2019deep,schmidt2020nonparametric,kohler2021rate}. Suppose that $f_0$ belongs to a hierarchical composition model with parameters $(p,d,\ell,C,\Pscr)$ and admits the representation $f_0=g_\ell\circ g_{\ell-1}\circ\cdots\circ g_1$, where $g_i:K_{i-1}\to K_i$ and $K_0=[0,1]^p$. Here, $K_0,\ldots,K_\ell$ are fixed bounded rectangles of compatible dimensions. Each coordinate function of $g_i$ depends on at most $t$ coordinates of its input and is $\alpha$-H\"older smooth with norm bounded by $C$, on the corresponding coordinate projection of $K_{i-1}$, for some $(\alpha,t)\in\Pscr$, where $\Pscr\subseteq[1,\infty)\times\NN_+$, $\sup_{(\alpha,t)\in\Pscr}(\alpha\lor t)<\infty$, and the component functions have uniformly bounded Lipschitz constants.

To exploit this compositional structure, we use the deep ReLU class~\eqref{eq:relu-entropic-class} for nonparametric OTRR. The following assumption specifies the data-generating process and the noise-density conditions needed for the convergence analysis.

\begin{assumption}\label{asum:relu-entropic-noise}
    Under model~\eqref{eq: data generating process of OT_rank regression}, suppose that $\xbf\in[0,1]^p$ and the noise $\xibf$ has finite second moment $\EE\|\xibf\|_2^2<\infty$ and a strictly positive density $p_{\xibf}\in C^2(\RR^d)$ satisfying $\nabla^2\{-\log p_{\xibf}(u)\}\preceq L_2 I_d$ for some $L_2>0$ and all $u\in\RR^d$. The target function $f_0$ belongs to the hierarchical composition model with parameters $(p,d,\ell,C,\Pscr)$ and satisfies $\EE\{f_0(\xbf)\}=0$ and $\|f_0\|_{2,L^\infty}\le L_1$. Define the effective complexity of this model by $\gamma^\ast:=\min_{(\alpha,t)\in\Pscr}\alpha/t$.
\end{assumption}

Assumption~\ref{asum:relu-entropic-noise} captures low-dimensional structure in the target function $f_0$ through the hierarchical composition model. It allows asymmetric heavy-tailed noise with finite second moments. To obtain the OT regularity needed for the analysis, the Hessian condition in Assumption~\ref{asum:relu-entropic-noise} bounds the curvature of the negative log-density from above. This condition is compatible with both log-concave and heavy-tailed noise distributions.

An explicit non-asymptotic convergence rate requires controlling the effect of OT-rank estimation explicitly. Unlike the linear-model analysis in Section~\ref{sec:full-linear-asymptotic-normality}, which remains agnostic about the choice of OT-rank estimator, the ReLU analysis requires uniform control of its effect on empirical risk increments over the candidate class. We therefore consider entropic OT-rank estimation in this section. Let $\nu$ be the uniform probability measure on $B_d(0,1)$, and let $u_n$ be the uniform probability measure on the sample labels $[n]$. For $f\in\Fcal_n$, write $R_i(f):=\ybf_i-f(\xbf_i)$ and, for $\tau>0$, define
\begin{equation}\label{eq:relu-entropic-plan}
    \pi_{n,\tau,f} :=\argmax_{\pi\in\Pi(u_n,\nu)} \left\{\sum_{i=1}^n\int R_i(f)^\top z\,\pi(i,\dd z) -\tau\operatorname{KL}(\pi\Vert u_n\otimes\nu)\right\}.
\end{equation}
Here $\operatorname{KL}$ denotes the Kullback--Leibler divergence. The entropic coupling is unique. Its conditional barycentres define the corresponding OT-rank estimator: $\widehat T_{n,\tau,f}\{R_i(f)\}:=n\int z\,\pi_{n,\tau,f}(i,\dd z)$. For $\tau_n>0$, define the empirical risk of OTRR with ReLU DNNs and entropic OT ranks by
\begin{equation}\label{eq:relu-entropic-estimator}
    \begin{aligned}
        \Lcal_{n,\tau_n}(f) &:=\frac1n\sum_{i=1}^n R_i(f)^\top\widehat T_{n,\tau_n,f}\{R_i(f)\},\qquad \widehat f_{n,\tau_n} \in\argmin_{f\in\Fcal_n}\Lcal_{n,\tau_n}(f).
    \end{aligned}
\end{equation}

Since OTRR identifies the regression function only up to an additive constant, we measure estimation error using the $L^2(P)$ distance modulo additive constants, defined by $d(f,g):=\inf_{c\in\RR^d}\|f-g-c\|_{2,L^2(P)}$. The following theorem bounds $d(\widehat f_{n,\tau_n},f_0)$ in terms of network complexity, approximation error, and entropic regularisation.

\begin{theorem}\label{thm:relu-entropic-oracle}
    Under Assumption~\ref{asum:relu-entropic-noise}, fix $p,d$, use the class~\eqref{eq:relu-entropic-class} with $D,W\ge3$, and let $\tau_n>0$ be deterministic. For every $n\ge2$ and $\delta\in(0,1/2)$, with probability at least $1-\delta$,
    \begin{equation}\label{eq:relu-entropic-oracle}
        d(\widehat f_{n,\tau_n},f_0) \lesssim \sqrt{\frac{H_n+\log(1/\delta)}{n}} +\inf_{f\in\Fcal_n}\|f-f_0\|_{2,L^2(P)} +\sqrt{\tau_n\log n},
    \end{equation}
    where $H_n:=\{1+(DW)^2\log(eDW)\}\log(en)$, and the suppressed constant may depend on $p_{\xibf}$, $p$, $d$, $L_1$ and $L_2$. In particular, choosing $D=D_n$ and $W=W_n$ with $D_n\asymp\log^2(en)$ and $W_n\asymp n^{\frac{1}{4\gamma^\ast+2}}\allowbreak\log(en)$  and $\tau_n\log n\lesssim n^{-\frac{2\gamma^\ast}{2\gamma^\ast+1}}$ then yields, with the same probability, $ d(\widehat f_{n,\tau_n},f_0)\lesssim_{\log} n^{-\frac{\gamma^\ast}{2\gamma^\ast+1}} + \sqrt{\frac{\log(1/\delta)}{n}}.$
\end{theorem}

With a suitable choice of the entropic regularisation parameter $\tau_n$, Theorem~\ref{thm:relu-entropic-oracle} establishes convergence in the $L^2(P)$ distance modulo additive constants, $d(\cdot,\cdot)$, at the minimax-optimal rate $n^{-\frac{\gamma^\ast}{2\gamma^\ast+1}}$ for hierarchical composition models~\cite{schmidt2020nonparametric}, up to logarithmic factors. Under the stated independent-noise and density regularity conditions, the exponent $\gamma^\ast/(2\gamma^\ast+1)$ reflects the target function's compositional structure and does not involve a noise-moment parameter. The bound also has sub-Gaussian dependence on the confidence parameter, thereby providing non-asymptotic robustness guarantees under heavy-tailed noise distributions.

\section{Numerical Studies}\label{sec: Numerical Studies}

This section evaluates the numerical performance of OTRR. Section~\ref{sec: Simulation Experiments} examines linear multivariate-response regression. Section~\ref{sec: Real Data Application} analyses NHANES oral-microbiome compositions. Nonparametric simulations and a complementary NHANES-based semi-synthetic experiment are presented in the Supplementary Material.

\subsection{Simulation Experiments for Linear Models}\label{sec: Simulation Experiments}

We consider $n$ samples from the linear model~\eqref{eq: data generating process of OT_rank regression for linear model}, allowing a nonzero covariate mean. Specifically, $\xbf_i\sim N(\boldsymbol\mu,I_p)$ with $\boldsymbol\mu=(1,1,1,1,0,\ldots,0)^\top$ and $(p,d)=(20,10)$. The randomly generated signal matrix $\Bbf_0$ has geometrically decaying singular values and is scaled so that $\|\Bbf_0\|_{\F}=\sqrt d$. We consider three asymmetric noise distributions. Noise A adds strong rank-one lognormal contamination to an isotropic Gaussian error, whereas Noise B uses rank-three Pareto contamination. Noise C is an asymmetric eight-component Gaussian mixture. The complete data-generating mechanisms are provided in the Supplementary Material.

For all three noise settings, we use training sample size $n=1000$ and compare LS, coordinatewise Huber (CW-Huber), Radial Huber, coordinatewise ASM (CW-ASM)~\cite{feng2026optimal}, coordinatewise CQR (CW-CQR)~\cite{zou2008composite}, MCQR with Gaussian reference measure~\cite{yang2024multiple}, and OTRR for \(50\) independent Monte Carlo repetitions. For OTRR, the candidate-dependent OT ranks are estimated using the entropic OT map estimator. Within each Monte Carlo repetition, we measure the coefficient error \(\|\widehat\Bbf-\Bbf_0\|_{\F}\). To assess orthogonal invariance empirically, we generate a random orthogonal matrix \mbox{\(Q\in\RR^{d\times d}\)} and fit each method to both the original sample \mbox{\(\{(\xbf_i,\ybf_i)\}_{i=1}^n\)} and the response-rotated sample \mbox{\(\{(\xbf_i,Q\ybf_i)\}_{i=1}^n\)}, obtaining \mbox{\(\widehat\Bbf\)} and \mbox{\(\widehat\Bbf_Q\)}, respectively. We map the rotated fit back to the original coordinates and report the maximum absolute coefficient difference: $E_{\max}:=\|Q^\top\widehat\Bbf_Q-\widehat\Bbf\|_{\max}$. Implementation details and corresponding side-by-side boxplots for Tables~\mbox{\ref{tab:simulation-pnw-current}} and~\mbox{\ref{tab:simulation-inference-current}} are provided in the Supplementary Material.

\begin{table}[!htb]
    \centering\setlength{\tabcolsep}{4pt}\renewcommand{\arraystretch}{1.2}
    \small
    \caption{Estimation accuracy and orthogonal invariance for linear models. Entries are Monte Carlo means, with standard deviations in parentheses. }
    \label{tab:simulation-pnw-current}
    \begin{adjustbox}{width=\textwidth}
    \begin{tabular}{l cc cc cc}
        \toprule
        & \multicolumn{2}{c}{Noise A} & \multicolumn{2}{c}{Noise B} & \multicolumn{2}{c}{Noise C}\\
        \cline{2-7}
        Method & Coef error & $E_{\max}$ & Coef error & $E_{\max}$ & Coef error & $E_{\max}$\\
        \midrule
        LS & 0.456 (0.056) & $<10^{-6}$ & 0.451 (0.042) & $<10^{-6}$ & 0.450 (0.067) & $<10^{-6}$\\
        Radial Huber & 0.242 (0.043) & $<10^{-6}$ & 0.244 (0.043) & $<10^{-6}$ & 0.175 (0.034) & $<10^{-6}$\\
        CW-Huber & 0.227 (0.045) & $0.02137\,(0.00607)$ & 0.231 (0.044) & $0.01933\,(0.00489)$ & 0.176 (0.035) & $0.01910\,(0.00654)$\\
        CW-ASM & 0.192 (0.047) & $0.01891\,(0.00501)$ & 0.198 (0.047) & $0.01844\,(0.00506)$ & 0.153 (0.033) & $0.01591\,(0.00507)$\\
        CW-CQR & 0.191 (0.047) & $0.01661\,(0.00434)$ & 0.195 (0.047) & $0.01615\,(0.00461)$ & 0.158 (0.040) & $0.01148\,(0.00336)$\\
        MCQR & 0.241 (0.057) & $0.05901\,(0.01660)$ & 0.252 (0.060) & $0.06343\,(0.01988)$ & 0.180 (0.039) & $0.04478\,(0.01182)$\\
        OTRR & \textbf{0.175 (0.043)} & $0.00032\,(0.00023)$ & \textbf{0.181 (0.043)} & $0.00030\,(0.00010)$ & \textbf{0.139 (0.034)} & $0.00024\,(0.00010)$\\
        \bottomrule
    \end{tabular}
    \end{adjustbox}
\end{table}

Table~\ref{tab:simulation-pnw-current} shows that OTRR has the smallest coefficient error in every setting and small $E_{\max}$. CW-ASM and CW-CQR improve on LS and Radial Huber in estimation accuracy but have larger $E_{\max}$ than OTRR. LS and Radial Huber are orthogonally equivariant by construction, but have larger coefficient errors than OTRR. CW-Huber is comparable to Radial Huber in estimation accuracy but is substantially more sensitive to response rotations. In these experiments, the implemented MCQR estimator has larger coefficient errors than CW-ASM and CW-CQR and the largest $E_{\max}$.

\begin{table}[!htb]
    \centering\setlength{\tabcolsep}{4pt}\renewcommand{\arraystretch}{1.2}
    \small
    \caption{Average coefficientwise CI lengths and coverage. Entries are Monte Carlo means, with standard deviations across repetitions in parentheses. }
    \label{tab:simulation-inference-current}
    \scalebox{0.8}{
    \begin{tabular}{llcccccc}
        \toprule
        Noise & Metric & LS & Radial Huber & CW-Huber & CW-ASM & CW-CQR & OTRR\\
        \midrule
        \multirow{4}{*}{A} & 95\% length & 0.108 (0.008) & 0.063 (0.010) & 0.061 (0.011) & 0.058 (0.010) & 0.052 (0.012) & \textbf{0.048 (0.011)}\\
        & 95\% cov. & 0.948 (0.030) & 0.945 (0.023) & 0.945 (0.023) & 0.962 (0.016) & 0.948 (0.019) & 0.948 (0.016)\\
        & 99\% length & 0.142 (0.011) & 0.083 (0.014) & 0.080 (0.014) & 0.076 (0.014) & 0.069 (0.015) & \textbf{0.063 (0.014)}\\
        & 99\% cov. & 0.992 (0.010) & 0.990 (0.010) & 0.990 (0.010) & 0.994 (0.006) & 0.990 (0.007) & 0.989 (0.007)\\
        \addlinespace
        \multirow{4}{*}{B} & 95\% length & 0.113 (0.008) & 0.065 (0.011) & 0.062 (0.012) & 0.059 (0.011) & 0.054 (0.012) & \textbf{0.049 (0.012)}\\
        & 95\% cov. & 0.954 (0.020) & 0.946 (0.020) & 0.946 (0.019) & 0.964 (0.012) & 0.949 (0.016) & 0.944 (0.017)\\
        & 99\% length & 0.149 (0.011) & 0.086 (0.015) & 0.082 (0.015) & 0.078 (0.015) & 0.070 (0.016) & \textbf{0.065 (0.015)}\\
        & 99\% cov. & 0.994 (0.007) & 0.988 (0.009) & 0.990 (0.007) & 0.995 (0.005) & 0.990 (0.007) & 0.989 (0.007)\\
        \addlinespace
        \multirow{4}{*}{C} & 95\% length & 0.108 (0.007) & 0.047 (0.008) & 0.048 (0.008) & 0.049 (0.009) & 0.043 (0.010) & \textbf{0.039 (0.009)}\\
        & 95\% cov. & 0.949 (0.033) & 0.950 (0.019) & 0.950 (0.020) & 0.972 (0.013) & 0.952 (0.016) & 0.956 (0.017)\\
        & 99\% length & 0.141 (0.009) & 0.062 (0.010) & 0.063 (0.011) & 0.065 (0.012) & 0.057 (0.013) & \textbf{0.051 (0.012)}\\
        & 99\% cov. & 0.991 (0.010) & 0.990 (0.007) & 0.990 (0.008) & 0.996 (0.004) & 0.989 (0.008) & 0.992 (0.007)\\
        \bottomrule
    \end{tabular}}
\end{table}

To compare inferential performance, for method $m$ and repetition $r$, let $I_{rj}^m(1-\alpha)$ denote the confidence interval (CI) for $\operatorname{vec}(\Bbf_0)_j$ at level $1-\alpha\in\{0.95,0.99\}$. We average lengths and coverage indicators over the $pd$ coefficients by $\frac{1}{pd}\sum_{j=1}^{pd}|I_{rj}^m(1-\alpha)|$ and $\frac{1}{pd}\sum_{j=1}^{pd}\II\{\operatorname{vec}(\Bbf_0)_j\in I_{rj}^m(1-\alpha)\}$, respectively. In Table~\ref{tab:simulation-inference-current}, OTRR yields the shortest average intervals with coverage close to the nominal levels, followed by CW-CQR; LS yields the longest average intervals. CW-ASM intervals are shorter than both Huber methods' under Noises A and B. MCQR is excluded because its asymptotic normality remains unestablished.

\subsection{Real Data Analysis: Oral Microbiome Composition}\label{sec: Real Data Application}

The oral microbiome is a complex microbial ecosystem associated with demographic, behavioural, health, and clinical characteristics~\cite{baker2024oral,chaturvedi2025oral}. However, weak signals, sparse data, compositionality, and high dimensionality make inference challenging~\cite{chaturvedi2025oral,nickols2026maaslin}. Moreover, many existing analyses examine microbial features separately~\cite{chaturvedi2025oral,xue2026population,nickols2026maaslin}, leaving substantial cross-taxon dependence unexploited~\cite{lin2022correlation}. These considerations motivate joint modelling to improve inferential efficiency.

Compositional responses also make orthogonal invariance essential for reliable inference: the orthonormal basis of an isometric log-ratio (ILR) representation is not unique, but inference for the same scientific contrasts should not depend on its choice. Relevant contrasts include the \textit{Actinomyces}-to-\textit{Haemophilus} ratio following tooth brushing~\cite{morton2019reference} and the \textit{Prevotella}-to-\textit{Rothia} and \textit{Prevotella}-to-\textit{Haemophilus} ratios associated with dental caries~\cite{baker2021deep}. Without orthogonal invariance, changing the ILR basis can alter their estimates, standard errors, and testing conclusions.

We analyse oral-microbiome profiles from 8,237 participants in the 2009--2012 U.S.\ National Health and Nutrition Examination Survey (NHANES), released in November 2024. Let $\mathbf c_i\in\Delta^7$ contain the relative abundances of \textit{Actinomyces}, \textit{Rothia}, \textit{Prevotella 7}, \textit{Gemella}, \textit{Streptococcus}, \textit{Veillonella}, \textit{Haemophilus}, and ``Other''. We use the ILR response $\ybf_i=H^\top\log(\mathbf c_i)\in\RR^7$, with $H\in\RR^{8\times7}$ the Helmert basis for $\mathbf1_8^\perp$. The 43 covariates cover demographic, socioeconomic, behavioural, metabolic, dietary, medication-use, and oral-health measures. Variable definitions and preprocessing are given in the Supplementary Material.

Using five-fold cross-validation, we fit linear models with LS, Radial Huber, CW-Huber, CW-ASM, CW-CQR, and OTRR. For out-of-fold predictions $\widehat\ybf_i^{\{-k(i)\}}$, where $k(i)$ is participant $i$'s fold, we report the mean absolute error $\operatorname{MAE}:=(nd)^{-1}\sum_i\|\ybf_i-\widehat\ybf_i^{\{-k(i)\}}\|_1$. Orthogonal invariance is assessed by $E_{\max}$ as in Section~\ref{sec: Simulation Experiments}, averaged over the five folds and two response rotations. For inference, we estimate genus-versus-Other effects by $\hat\theta_{gj}:=(\ebf_g-\ebf_{\mathrm{Other}})^\top H\widehat\Bbf_n\ebf_j$, $g\in[7]$, $j\in[43]$, where the $\ebf$ vectors select the indicated components. We report their average 95\% confidence-interval length under the Helmert representation. For each method, we form selected contrast sets under the Helmert basis and two orthogonal rotations. Selection requires a common sign and a Benjamini--Hochberg (BH) adjusted $p<0.05$~\cite{benjamini1995controlling} in all five folds, adjusting jointly over the $43\times7$ contrasts within each fold. The significant-contrast intersection over union (IoU) is the intersection size of the three sets divided by their union size.

Table~\ref{tab:real-data-nhanes} reports these four measures. It shows nearly identical MAE across methods, consistent with the limited variation explained by measured participant characteristics in previous research~\cite{chaturvedi2025oral}. LS and Radial Huber are orthogonally invariant by construction; OTRR has mean $E_{\max}=9.80\times10^{-4}$, compared with $0.0167$--$0.0292$ for the coordinatewise methods. OTRR also yields the shortest CIs.

\begin{table}[!htb]
    \centering\setlength{\tabcolsep}{4pt}\renewcommand{\arraystretch}{1.2}
    \small
    \caption{NHANES oral-microbiome results.  Parentheses report bootstrap SDs from resampling participants for MAE and folds for mean CI length, and sample SDs across ten fold--rotation pairs for $E_{\max}$.}
    \label{tab:real-data-nhanes}
    \begin{adjustbox}{max width=0.8\textwidth}
    \begin{tabular}{lcccc}
        \toprule
        Method & MAE & $E_{\max}$ & 95\% CI length & IoU\\
        \midrule
        LS & 0.5954 (0.0035) & $<10^{-6}$ & 0.1393 (0.0002) & 48/48\\
        Radial Huber & 0.5948 (0.0035) & $<10^{-6}$ & 0.1294 (0.0002) & 56/56\\
        CW-Huber & \textbf{0.5943 (0.0036)} & $0.01672\,(0.00375)$ & 0.1197 (0.0001) & 50/70\\
        CW-ASM & 0.5950 (0.0036) & $0.02923\,(0.00506)$ & 0.1171 (0.0002) & 54/76\\
        CW-CQR & 0.5944 (0.0036) & $0.01732\,(0.00189)$ & 0.1167 (0.0001) & 53/74\\
        OTRR & \textbf{0.5943 (0.0036)}  & $0.00098\,(0.00040)$ & \textbf{0.1137 (0.0001)} & \textbf{71/72}\\
        \bottomrule
    \end{tabular}
    \end{adjustbox}
\end{table}

Moreover, OTRR yields substantially more basis-independent inferential conclusions. Across the three orthonormal representations, 71 of the 72 contrasts selected by OTRR are selected under every representation. LS and Radial Huber retain all their discoveries but identify only 48 and 56 contrasts, respectively. Coordinatewise methods identify approximately 70 contrasts in total, but retain only about 70\% of their respective unions across representations, revealing substantial basis dependence. Because the true signals in NHANES are unknown, a complementary semi-synthetic experiment uses its predictor design with known effects to show how basis dependence can reduce testing power. See Section~\ref*{sec: Implementation Details of Oral Microbiome Analysis} of the Supplementary Material for this semi-synthetic experiment.

Lastly, we highlight three OTRR contrasts that are stable across all three response representations but are not selected by any baseline method and are directionally consistent with previous empirical findings. OTRR associates increasing age with a lower \textit{Streptococcus}-to-Other log-ratio, consistent with the lower abundance of early colonisers such as \textit{Streptococcus} observed in older adults~\cite{park2026age}. The positive associations of glycated haemoglobin with the \textit{Prevotella 7}-to-Other and \textit{Streptococcus}-to-Other log-ratios are aligned with evidence from an oral-microbiome study linking \textit{Prevotella micans} and \textit{Streptococcus parasanguinis} to higher glucose levels or glycaemic variability~\cite{xue2026population}.

\section{Conclusion}\label{sec: Conclusion}

This paper develops OTRR for robust multivariate-response mean regression by exploiting cross-response dependence to improve efficiency. We establish Fisher consistency and orthogonal invariance as intrinsic properties of its population risk. For linear models, we derive a $\sqrt{n}$-Gaussian limit, establish bounded asymptotic influence conditional on the covariates, and develop asymptotically basis-independent Wald inference. We also bound its asymptotic efficiency relative to LS from below, uniformly over noise distributions, and compare OTRR with likelihood-based estimators. For deep ReLU networks, we show that OTRR attains the minimax-optimal rate in $L^2(P)$ modulo additive constants, up to logarithmic factors, and we provide non-asymptotic robustness guarantees. Simulations for linear models show gains in estimation accuracy and inferential efficiency under asymmetric noise, including heavy-tailed, contaminated, and mixture designs, with the estimator remaining nearly orthogonally invariant in finite samples. The NHANES analysis highlights the importance of orthogonal invariance for stable scientific discoveries, and the accompanying semi-synthetic experiment illustrates its implications for testing power. These findings support OTRR as a theoretically grounded and practically useful approach to robust multivariate-response mean regression.

\printbibliography

\end{document}